\documentclass[prx,twocolumn,longbibliography,superscriptaddress]{revtex4-1}

\usepackage{epsfig,amsmath,amssymb}
\usepackage[outdir=./]{epstopdf}
\usepackage{xcolor}

\newcommand{\bra}[1]{\langle #1|}
\newcommand{\ket}[1]{|#1\rangle}
\newcommand{\Exp}[1]{\langle#1\rangle}
\newcommand{\Lim}[1]{\raisebox{0.5ex}{\scalebox{0.8}{$\displaystyle \lim_{#1}\;$}}}

\begin{document}
\title{Using dark states to charge and stabilize open quantum batteries} 
%\title{Using dark states to stably charge open quantum batteries \\ Charging open quantum batteries with dark states \\ Anti-thermalization as a stable charging mechanism for open quantum batteries} 

\author{James Q. Quach} 
\email{quach.james@gmail.com}
\affiliation{Institute for Photonics and Advanced Sensing and School of Chemistry and Physics, The University of Adelaide, South Australia 5005, Australia}

\author{William J. Munro} 
\affiliation{NTT Basic Research Laboratories \& NTT Research Center for Theoretical Quantum Physics, NTT Corporation, 3-1 Morinosato-Wakamiya, Atsugi-shi, Kanagawa 243-0198, Japan}
\affiliation{National Institute of Informatics, 2-1-2 Hitotsubashi, Chiyoda-ku, Tokyo 101-8430, Japan}

\begin{abstract}
We introduce an open quantum battery protocol using dark states to achieve both superextensive capacity and power density, with non-interacting spins coupled to a reservoir. Further, our power density actually scales with the of number of spins $N$ in the battery. We show that the enhanced capacity and power is correlated with entanglement. Whilst connected to the charger, the charged state of the battery is a steady state, stabilized through quantum interference in the open system.
\end{abstract}

\maketitle
\section{Introduction}
\label{sec:Intro}

The recent interest in quantum technologies is driven by the potential power of quantum mechanics~\cite{nielsen02,dowling03,spiller05}, and the push towards technological miniaturization Harnessing the unique properties of quantum mechanics, such as entanglement and superposition, promises to open new vistas in computing, sensing, cryptography, and other quantum technologies~\cite{feynman82,caves82,bennett84,deutsch85,lloyd96,bennet96,vanenk98,shor99,gisin02,giovannetti04,gisin07,ladd10}. The increasing rate of technology miniaturization, in particular electronics, has meant that we need to account for quantum effects. This has driven the relatively new field of quantum thermodynamics, which tries to understand thermodynamic concepts such as work, heat, and entropy in a quantum context~\cite{kosloff13,pekola15,vinjanampathy16,goold16,anders17,lostaglio19,binder19,mitchison19}. Quantum batteries (QBs) aim to harness the unique properties of quantum thermodynamics to build batteries that are fundamentally different from conventional batteries~\cite{alicki13,hovhannisyan13,skrzypczyk14,binder15}.  

Typically, QBs were  modeled as a collection of $N$ identical quantum subsystems to which an external field, which acted as the energy source, was applied~\cite{alicki13}. Alicki and Fannes~\cite{alicki13} sought to understand whether entanglement could enhance the amount of extractable work in this model. Under closed unitary evolution, they showed that one can extract more work with entanglement than without. Further work revealed that it may be possible to reduce the amount of entanglement without detrimentally affecting the maximal work extraction, with the caveat that with reduced entanglement one requires more operations~\cite{hovhannisyan13}. This then lead to the notion that entanglement boosted the charging rate of QBs, as it reduced that number of traversed states in the Hilbert space between the initial and final separable states~\cite{hovhannisyan13}. This conjecture was  supported by Binder \textit{et al.}~\cite{binder15}, who showed that entangled spins can superextensively charge $N$ time faster than non-interacting spins, where $N$ is the number of spins. The main finding was that using global entangling operators, where all spins can interact with each other, can result in a speed-up of the charging power as compared to charging them individually. Further work argued that \textit{N} power scaling is the theoretical upper bound of the quantum advantage, constrained by quantum speed limits~\cite{campaioli17}

All these studies assumed global operators, which in practice is difficult to implement. Ferraro \textit{et al.}~\cite{ferraro18} overcame this problem by showing that, by locally coupling all of the spins coherently to the same quantum energy source in a photonic cavity, one can realize effective long-range interactions amongst all the spins. Known as the Dicke QB, after the Hamiltonian that describes it, they showed that the time taken to reach the maximum stored energy in the spin ensemble reduced as the ensemble got larger, such that the charging power scaled with $\sqrt{N}$ for large $N$. This increased the potential for QBs to be physically realized. However, recent work has shown that entanglement does not underlie the charging speedup in the Dicke QB, instead it is the result of an enhanced effective cavity coupling strength, which arises out of coherent cooperative interactions~\cite{zhang18}.

Recently, QBs have been considered in an open system context~\cite{farina19,liu19}. This is important as QB must interact with its environment for the device to ever be practical. In particular, protocols are needed to stabilize the charged state of the QB in an open system. A recent attempt proposed the continual measurement of the system for stabilization~\cite{gherardini19}. However, this protocol requires continuous access to the battery, and the measurement process itself is costly, consuming energy.

Here we use dark states to achieve both superextensive capacity and power, that scales with $N$, with only local interactions, in an open system.  We will show that the superextensive behavior of the system is correlated with entanglement. Furthermore, the stored energy of the battery is stable without the need to continually access the battery.

\section{Model} 
\label{sec:Model} 
\begin{figure}[tb]
	\includegraphics[width=.5\columnwidth]{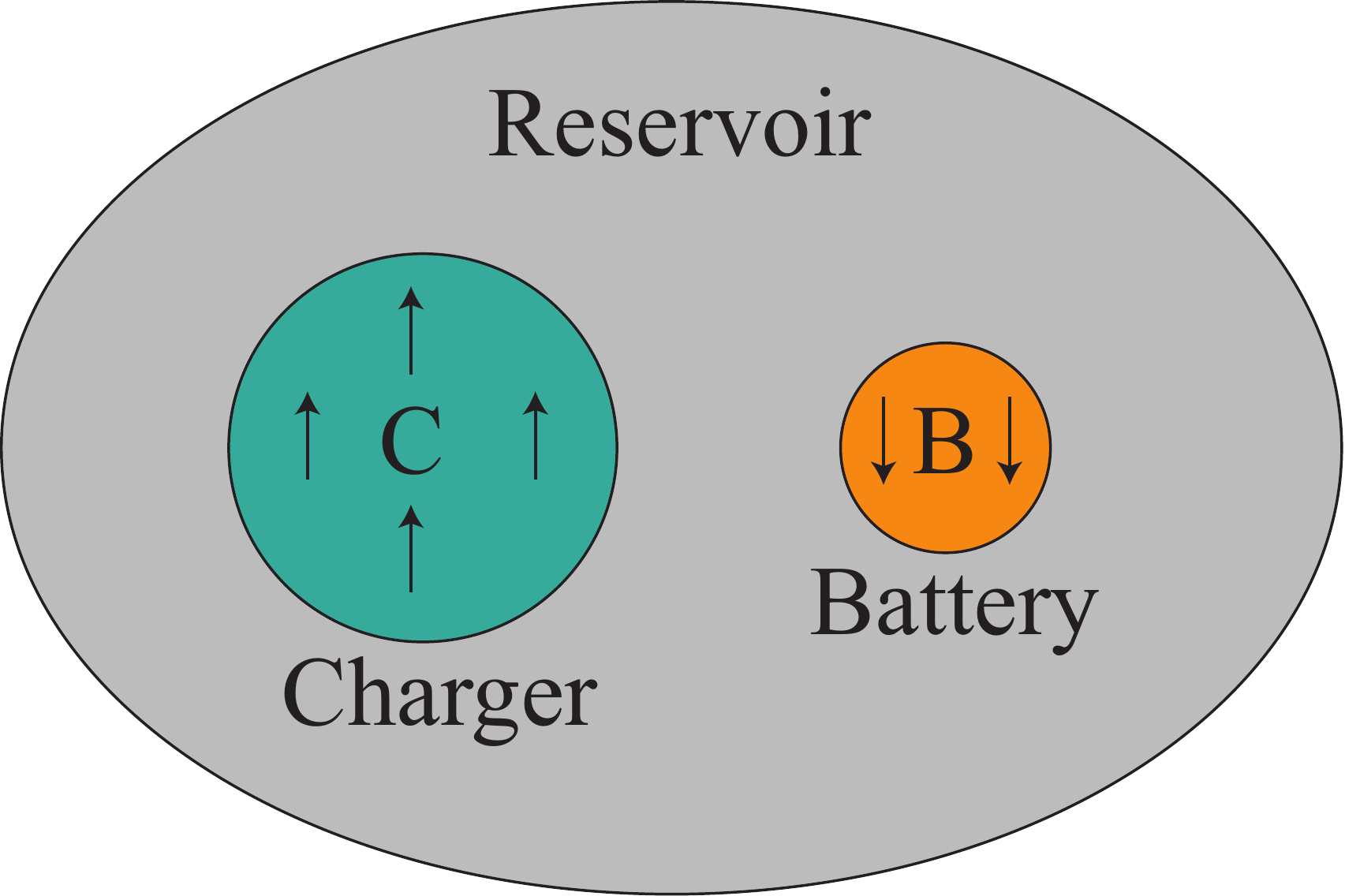}
	\caption{\textit{Model.} The spin-charged QB is modeled as an ensemble of spins in a reservoir. Initially, the QB is in thermal equilibrium with the reservoir. The charger is another ensemble of spins, but in the excited state. The charging process is initiated by bringing the charger into the reservoir.}
	\label{fig:1}
\end{figure}

In general, the QB charging protocol consists of a battery and an energy source or charger. Switching on (off) the coupling between the battery and charger initiates the charging (discharging) process. We consider a QB in an open system, modeled as an ensemble of $N_B$ $\frac12$-spins with transition energy $\hbar\omega$, in a thermal reservoir Fig.~\ref{fig:1}. Initially, the QB is in thermal equilibrium with the reservoir. The charger is another ensemble of $N_C$ $\frac12$-spins, but in the excited (up) state. We will assume $N_C\ge N_B$. The charging process is initiated by bringing the charger into the reservoir. 

The Hamiltonian of our model is
\begin{equation}
\begin{split}
H &= \omega(J_B^z+J_C^z)+\int d^dkE_\mathbf{k}r_\mathbf{k}^\dagger\\
&+\frac{g}{2}[(J_B^++J_C^+)R+(J_B^-+J_C^-)R^\dagger]
\end{split}
\label{eq:H}
\end{equation}
where  $J^{x,y,z}_i$ are the usual collective spin operators on ensemble $i$, with the collective raising and lowering operators defined as $J^\pm_i=J^x_i\pm iJ^y_i$. The first of term of the Hamiltonian represents the battery and charger. The second term represents the reservoir with $d$ spatial dimension and wave vectors $\mathbf{k}=(k_1,\cdots,k_d)$. $E_\mathbf{k}$ is the linear dispersion relation with $r_\mathbf{k}(r_\mathbf{k}^\dagger)$ the annihilation (creation) operator satisfying the commutation relation $[r_\mathbf{k},r_\mathbf{k}^\dagger]=\delta(\mathbf{k}-\mathbf{k}')$. The third term is the interaction between this reservoir and the spins with coupling strength $g$, where $R=\int d^dk\kappa_\mathbf{k}r_\mathbf{k}$ with $\kappa_\mathbf{k}$ being a continuous function of $\mathbf{k}$ whose exact form depends on the system under consideration.

Under the assumption that the spins and reservoir were initially uncorrelated, one can characterize the reservoir with the density matrix $\rho_R=\exp(-H_R/k_BT)/\text{Tr}_R[\exp(-H_R/k_BT)]$ where $H_R=\int d^dkE_\mathbf{k}r_\mathbf{k}^\dagger$. With the Born-Markov approximation, the Lindblad master equation during charging is~\cite{hama18},
\begin{equation}
\begin{split}
\dot{\rho}(t)&=-i\frac{\omega}{\hbar}[J_C^z+J_B^z,\rho(t)]\\
&\quad+\frac{\gamma}{\hbar^2}\Big[(\bar{n}+1)\mathcal{L}(J_C^-+J_B^-)+\bar{n}\mathcal{L}(J_C^++J_B^+))\Big]~,
\end{split}
\label{eq:me}
\end{equation}
where $\mathcal{L}(O)\equiv2O\rho O^\dagger-O^\dagger O\rho-\rho O^\dagger O$ is the Lindblad superoperator. The damping rate $\gamma$ is a function of  $g$ and $|\kappa_\mathbf{k}|^2$, and $\bar{n}=1/(e^{\hbar\omega/k_BT}-1)$ is the mean thermal population. Importantly, even though we have a non-interacting spin model, the Lindblad operator gives rise to terms that affects a global spin entangling operator.

\section{Energy transfer and stabilization mechanism} 
\label{sec:Energy transfer and stabilisation mechanism}
Naively, one may expect all energy to be loss to the reservoir at zero temperature. However, quantum interference can lead to steady states that are not the ground state. Consider the two spin case at $T=0$, which at initial time is 
$\ket{\psi_0}=\ket{\frac12}_C\ket{{-\frac12}}_B$. This can be expressed as
\begin{equation}
\rho_0 =\frac12(\ket{\psi_+}\bra{\psi_+}+\ket{\psi_+}\bra{\psi_-}+\ket{\psi_-}\bra{\psi_+}+\ket{\psi_-}\bra{\psi_-})~,
\end{equation}
where $\ket{\psi_\pm}\equiv(\ket{\frac12}_C\ket{{-\frac12}}_B\pm\ket{{-\frac12}}_C\ket{\frac12}_B)/\sqrt{2}$~. The anti-symmetric component does not couple to the reservoir, since $\mathcal{L}(J_C^-+J_B^-)=0$ for $\ket{\psi_-}\bra{\psi_-}$, and therefore does not decay. Such states are known as dark or subradiant states~\cite{freedhoff67,stroud72}. The other components decay to the ground state leading to a steady state of the form
\begin{equation}
	\rho^\text{ss}=\frac{1}{2}\ket{\psi_\downarrow}\bra{\psi_\downarrow}+\frac{1}{2}\ket{\psi_-}\bra{\psi_-}~,
\label{eq:rho_ss}
\end{equation}
where $\ket{\psi_\downarrow}\equiv\ket{{-\frac12}}_C\ket{{-\frac12}}_B$ (see Appendix for a formal derivation). In this steady state the spin angular momenta of the charger and battery are 
\begin{equation}
	\Exp{J_C^z}=-\frac{\hbar}{4}~,\quad\Exp{J_B^z}=-\frac{\hbar}{4}~,
\end{equation}
where $\Exp{J_i^z}=\mathrm{Tr}(\rho_iJ_i^z)$. We immediately observe that $\hbar\omega/4$ units of energy has been transferred from the charger to the battery, since initially $\Exp{J_B^z}=-\hbar/2$. One notes that this transfer of energy cannot be viewed (semi-) classically as a transfer of energy due to the emission of a photon by the charger followed by the absorption of that photon by the battery. Instead, this is a purely quantum mechanical effect which arises out of the the collective behavior of the battery, charger, and reservoir. As the steady state is decoupled from the environment, the stored energy of the battery is stable whilst the charger is present, even in the open system. This is the basis of how energy is transferred and stably stored in our open system protocol. In general, for this effect to take place the initial combined battery and charger states should overlap with a dark state(s). This condition is trivially satisfied when the charger state is initially excited and the battery is in its ground state. One notes that dark states have been proposed to stabilize energy storage in a single three-level system~\cite{santos19}; what we are proposing here is very different, involving the collective effect of multiple two-level systems.

\begin{figure*}[tb]
	\includegraphics[width=1.75\columnwidth]{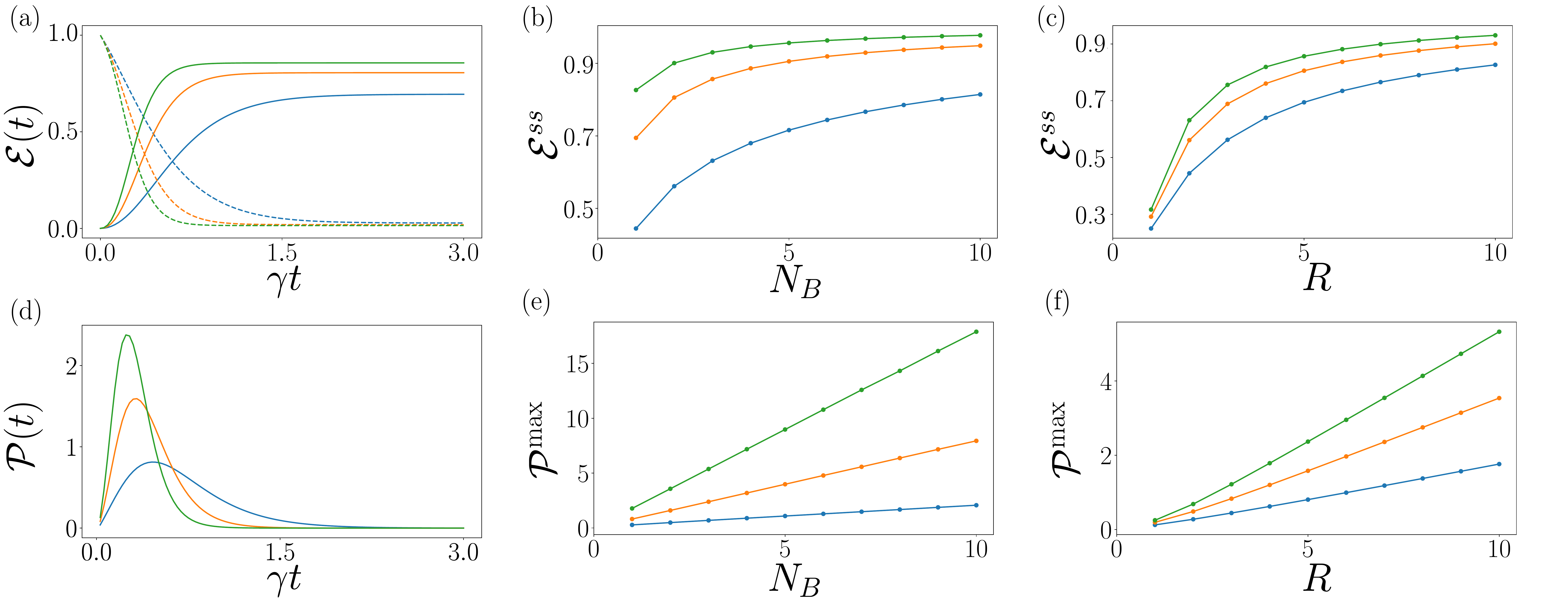}
	\caption{\textit{Superextensive capacity and power density.} (a) The energy density of the charger $\mathcal{E}_C(t)$ (dotted) and battery $\mathcal{E}_B(t)$ (solid) during the charging process. $\mathcal{E}_C(t)$ decreases as $\mathcal{E}_{B}(t)$ correspondingly increases, indicating a transfer of energy from charger to the battery. (b) The monotonic increase of the steady state energy density $\mathcal{E}^\text{ss}_B$ with $N_B$, shows the superextensive increase in battery capacity. (c) $\mathcal{E}^\text{ss}_B$ monotonically increases with $R$.	(d) The power density of the battery $\mathcal{P}_B(t)$ during the charging process.  (e) Peak power density $\mathcal{P}_B^\text{max}$ superextensively scales with $N_B$. (f) $\mathcal{P}_B^\text{max}$ monotonically increases with $R$. \textit{Parameters}: (a),(d) R = 5,  $N_B$ = 1 (blue), 2 (orange), 3 (green). (b),(e) R = 2 (blue), 5 (orange), 10 (green). (c),(f) $N_B$ = 1 (blue), 2 (orange), 3 (green). $\mathcal{E}$ and $\mathcal{P}$ are in units of $\hbar\omega$, with dimensionless $\gamma t$.}
	\label{fig:2}
\end{figure*}

\section{Superextensive capacity and charging}
\label{sec:Superextensive capacity}
\textit{Superextensive capacity.} The energy density of the charger and battery are
\begin{equation}
	\mathcal{E}_i(t)=\omega\Big(\frac{\Exp{J^z_i(t)}}{N_i}+\frac12\Big)~,
\end{equation}
with $i=B,C$. The capacity of the battery is defined as the energy in the steady state,
\begin{equation}
	E_{R,N_B} \equiv N_B\mathcal{E}_B^\mathrm{ss}~,
\end{equation}
where $\mathcal{E}^\text{ss}_B$ is the steady state energy density with $R\equiv N_C/N_B$ being the ratio of the number of spins in the charger to the battery. We have shown for the case where $N_C=N_B=1$, that the steady state energy of the battery is $\hbar\omega/4$. If we had $M$ of these systems isolated from each other, the energy density would not change, so that the total capacity would be $E_{1,M}=M\mathcal{E}_B^{ss}=M\hbar\omega/4$. However, we can improve on this by charging the batteries collectively. 

As a example, let us consider the case with $R=5$ during charging. Solving the master equation [Eq.~(\ref{eq:me})] at zero temperature, we plot $\mathcal{E}_i(t)$ for $N_B=1,2,3$ in Fig.~2(a). Firstly, the plots show that $\mathcal{E}_C(t)$ monotonically decreases as $\mathcal{E}_{B}(t)$ correspondingly increases, indicating a transfer of energy from charger to the battery. Secondly, $\mathcal{E}^\text{ss}_B$ increases with $N_B$. This is shown in Fig.~\ref{fig:2}(b) where we plot $\mathcal{E}^\text{ss}_B(N_B)$ for $R=2,5,10$. As $\mathcal{E}^\text{ss}_B(N_B)$ increases monotonically, the capacity of the battery scales superextenstively. With increasing $R$, the scaling of $\mathcal{E}^\text{ss}_B$ with $N_B$ decreases, \textit{i.e.} the plot tends to flatten out, even for small $N_B$. This indicates a decrease in the superextensive capacity of the battery with $R$. Fig.~\ref{fig:2}(c) plots $\mathcal{E}^\text{ss}_B(R)$ for $N_B=1,2,3$. In the thermodynamic limits, $\Lim{R \rightarrow \infty} \mathcal{E}^\text{ss} = \Lim{N_B \rightarrow \infty} \mathcal{E}^\text{ss} = \hbar\omega/2~,\forall R>1$~. 

The superextensive scaling of $\mathcal{E}^\mathrm{ss}$ means that the capacity of one battery with $M$ spins is greater than $M$ batteries with one spin, \textit{i.e.} $E_{R,M}>ME_{R,1}~,~\forall M>1$. This improves upon the Dicke QB, where the capacity does not in general superextensively scale with the number of spins~\cite{ferraro18}. 
%As entanglement underpins the superextensive capacity of the battery, our protocol is a realization of the entanglement energy boost envisioned by Alicki and Fannes~\cite{alicki13}. However, unlike their closed global interaction model, our protocol is an open local interaction model, and therefore is more practical to realize.

\textit{Ergotropy.} One notes that not all stored energy may be extractable as work. In an open system, the thermal state energy ($\mathcal{E}_B^\text{th}$) represents a natural limit on extractable work as
\begin{equation}
	\mathcal{W}^\text{open}=\mathcal{E}_B-\mathcal{E}_B^\text{th}~.
\end{equation} 
For zero temperature $\mathcal{E}_B^\text{th}=0$, and so $\mathcal{W}^\text{open}=\mathcal{E}_B$. Another class of extractable work occurs under unitary evolution of the battery, and is known as \textit{ergotropy}. The ergotropy of a system is the maximal amount of work that can be extracted acting cyclically under thermal isolation. This is an important measure, as not all the energy stored in a system can be unitarily extracted as work. The ergotropy density is given by~\cite{allahverdyan04}
\begin{equation}
\mathcal{W}^\text{closed}=\mathcal{E}_B-\min_{U_B}\mathcal{E}_B~,
\end{equation} 
where the second term is the minimum battery energy under all possible unitary evolution of the battery $U_B$. The $\min_{U_B}\mathcal{E}_B =\omega\min_{U_B} \text{Tr}(J^z_BU_B\rho_BU^\dagger_B)/N_B$ term can be found by ordering the eigenvalues of $J_B^z/N_B$ in increasing order ($\epsilon_1<\epsilon_2<\cdots<\epsilon_n$), and the eigenvalues of $\rho_B$ in decreasing order ($r_1<r_2<\cdots<r_n$). From this we get that~\cite{allahverdyan04}
\begin{equation}
\min_{U_B}\mathcal{E}_B =\omega\sum_ir_i\epsilon_i~.
\end{equation}

\begin{figure}[hbt]
	\includegraphics[width=\columnwidth]{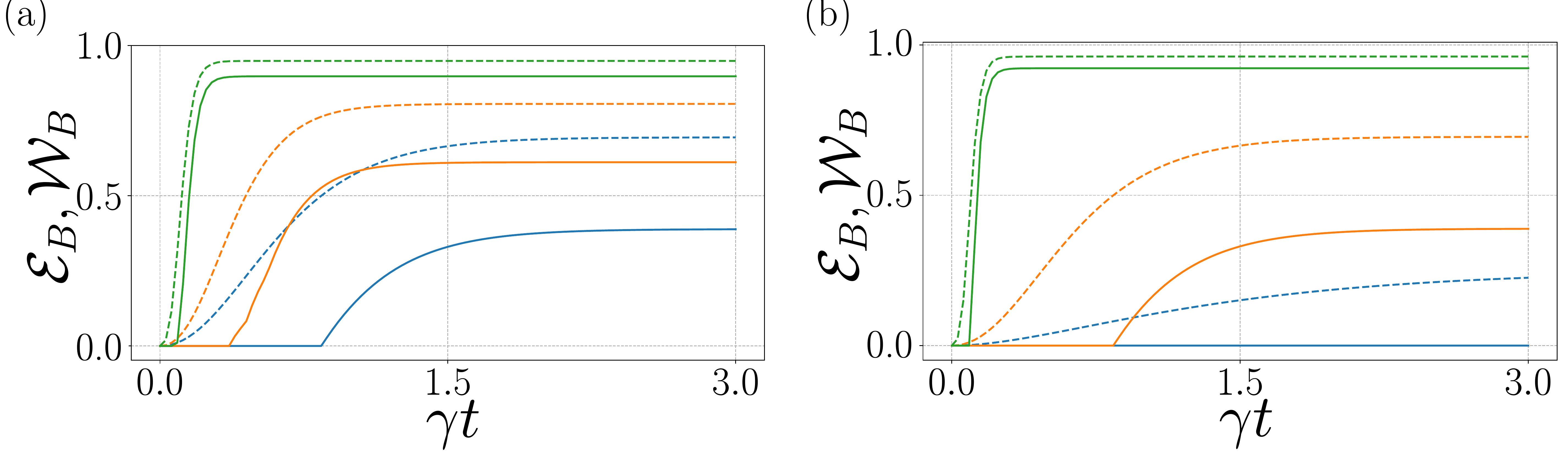}
	\caption{Ergotropy (solid line) and stored energy (dotted line) for (a) various $N_B$ and $R=5$, (b) $R$ and $N_B=1$. For $N_B=1$ the ergotropy is zero until $\mathcal{E}_B>\frac12$. As $N_B$ increases, the ergotropy approaches the stored energy. For $R=1$ the ergotropy is always zero, as the stored energy is always negative. As $R$ increases, the ergotropy approaches the stored energy. \textit{Parameters}: (a) $N_B$ = 1 (blue), 2 (orange), 10 (green); (b) $R$ = 1 (blue), 5 (orange), 50 (green). The vertical axes are in units of $\hbar\omega$. }
	\label{fig:A1}
\end{figure}

It is conjectured that $\mathcal{W}^\text{closed}\rightarrow\mathcal{E}_B$ in the large $N_B$ limit~\cite{andolina19}. This is a particular useful conjecture as this would mean that in principle nearly all the stored energy could be extracted as work, in most practical applications. Our system is indeed consistent with this conjecture. In addition, we find that $\mathcal{W}^\text{closed}\rightarrow\mathcal{E}_B$ in the large $R$ limit also. We plot in Fig.~\ref{fig:A1}(a) the ergotropy for $R=5$ for various $N_B$. For $N_B=1$ the ergotropy is zero until the stored energy $\mathcal{E}_B>\frac12$ (or $\Exp{J_B^z}>0$). As $N_B$ increases, the ergotropy approaches the stored energy. Fig.~\ref{fig:A1}(b) plots the ergotropy for $N_B=1$ various $R$. As $R$ increases, the ergotropy approaches the stored energy. The figures shows that work can only be extracted in a cyclic manner when there is a net positive spin angular moment, $\Exp{J_B^z}>0$.

%\section{Superextensive charging}  
\label{sec:Superextensive charging}
\textit{Superextensive charging.} The power density of the battery is given by
\begin{equation}
	\mathcal{P}_B(t) = \frac{d\mathcal{E}_B(t)}{dt}~,
\end{equation}
which we plot in Fig.~\ref{fig:2}(d). The plot shows that maximum power density $\mathcal{P}_B^\text{max}$ increases with $N_B$. This is clearly shown in Fig.~2(e) where we observe that $\mathcal{P}_B^\text{max}(N_B)\propto N_B$. Up until now, charging protocols have required global interactions to achieve $N$ scaling~\cite{hovhannisyan13,binder15,le18}. Protocols with local interactions have not exceeded $\sqrt{N}$ scaling~\cite{ferraro18,le18,zhang18}. Here we have shown that one can achieve $N$ power scaling with non-interacting spins coupled to a reservoir. Fig.~\ref{fig:2}(f) shows that $\mathcal{P}_B^\text{max}(R)$ also scales with $R$.

As the battery superextensively charges, if one were to simply disconnect the charger, it would also superextensively discharge as well. The reason for this is that the coherent spins would superradiantly decay~\cite{dicke54}.  However, if a slow discharge is desired, we propose an intermediately process of dephasing to destroy spin coherence, before disconnecting the charger. This could be achieved with a dephasing pulse, for example. With no coherence, the battery would discharge at the single-spin relaxation rate.

\section{Entanglement}
\label{sec:Entanglement}
\begin{figure}[tb]
\includegraphics[width=\columnwidth]{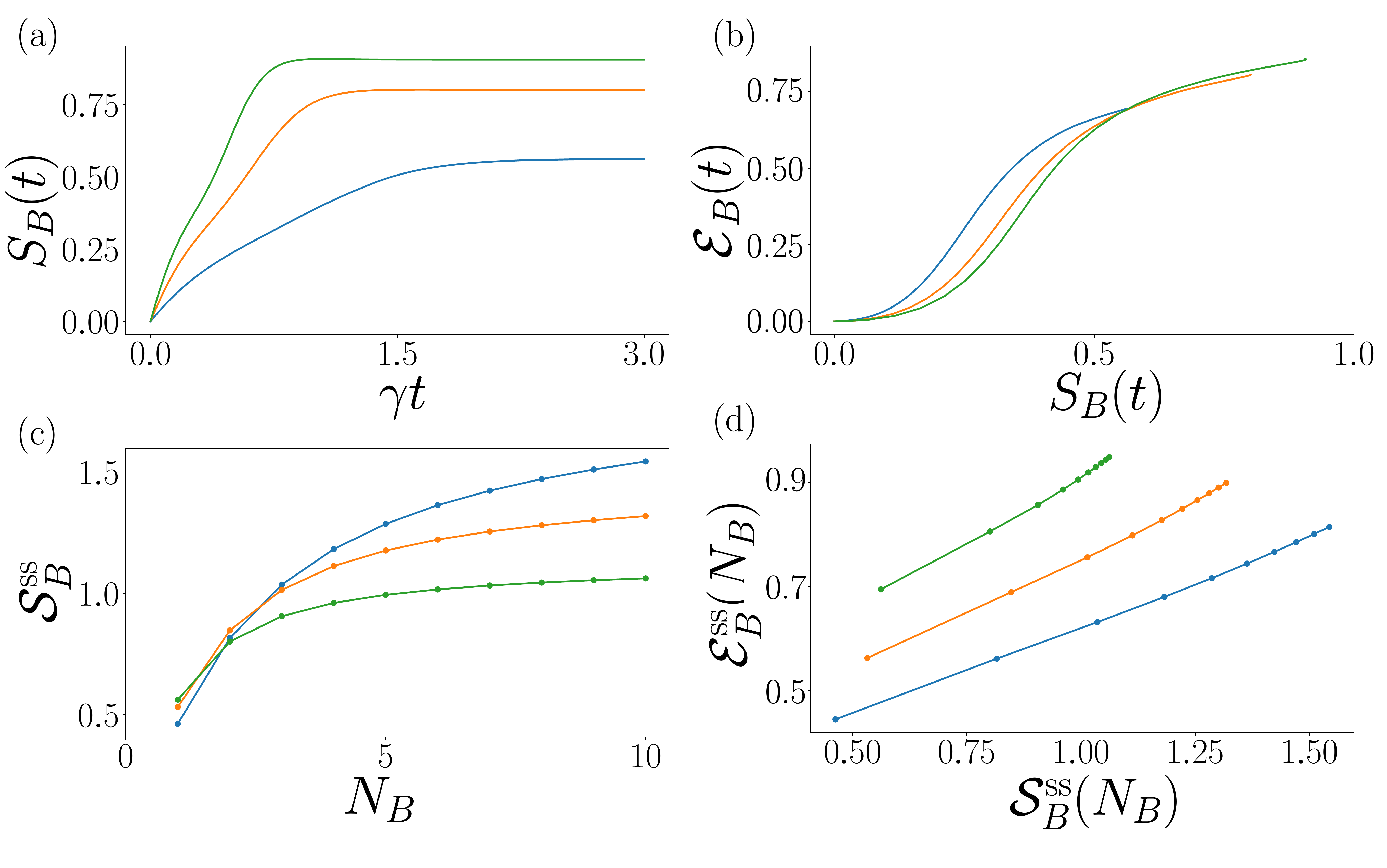}
	\caption{\textit{Entanglement and capacity.} (a) Logarithmic negativity $S_B(t)$. Comparing this plot with $\mathcal{E}_B(t)$  in Fig.~\ref{fig:2}(a), shows that higher entanglement corresponds to higher energy density.  (b) The relationship between entanglement and energy density is shown in this parameterized plot of $\mathcal{E}_B(t)$ and $\mathcal{S}_B(t)$. (c) $\mathcal{S}_B^\mathrm{ss}$ scales positively with $N_B$. Comparing this plot with $\mathcal{E}^\mathrm{ss}_B(N_B)$  in Fig.~\ref{fig:2}(b), shows that higher entanglement corresponds to higher energy density, in the steady state.  (d) $\mathcal{E}_B^\mathrm{ss}$ and $\mathcal{S}_B^\mathrm{ss}$ parameterized over $N_B$, shows the positive correlation between the capacity and entanglement. \textit{Parameters}: (a),(b) R = 5, $N_B$ = 1 (blue), 2 (orange), 3 (green). (c),(d) R = 2 (blue) , 3 (orange), 5 (green).}
	\label{fig:3}
\end{figure}

The role of entanglement has been studied in closed unitary QB systems~\cite{alicki13,hovhannisyan13,binder15,le18,zhang18}. Here we systematically investigate the role of entanglement in our open QB protocol.  For mixed systems, the logarithmic negativity~\cite{vidal02,plenio05} provides a convenient measure of entanglement. It is defined using the trace norm as
\begin{equation}
	S_B(t) = \log_2\|\rho^{\Gamma_B}(t)\|~,
\end{equation}
where $\Gamma_B$ denotes the partial transpose with respect to subsystem $B$. We plot $S_B(t)$ in Fig.~\ref{fig:3}(a), with the same parameters as Fig.~\ref{fig:2}(a). A comparison of these two plots shows higher entanglement to correspond to higher energy, supporting the idea that entanglement drives the superextensive capacity of the battery. Their relationship is shown in Fig.~\ref{fig:3}(b), where we plot $\mathcal{E}_B(t)$ and $\mathcal{S}_B(t)$ parameterized over $t$. In Fig.~\ref{fig:3}(c), we plot $\mathcal{S}_B^\mathrm{ss}(N_B)$, showing that steady state entanglement scales positively with $N_B$. In Fig.~\ref{fig:3}(d) we plot $\mathcal{E}_B^\mathrm{ss}(N_B)$ and $\mathcal{S}_B^\mathrm{ss}(N_B)$ parameterized over $N_B$, showing the positive correlation between the battery capacity and entanglement.

Revealingly, Fig.~\ref{fig:3}(d) shows that entanglement decreases with increasing $R$ (for a given $\mathcal{E}^\text{ss}_B$), inline with the decreased superextensive scaling of $\mathcal{S}_B^\mathrm{ss}$ in Fig.~\ref{fig:2}(b). In other words, as $R$ increases we have less entanglement to drive the system, and hence the ability of the battery capacity to superextensively increase, diminishes.

\begin{figure}[tb]
	\includegraphics[width=\columnwidth]{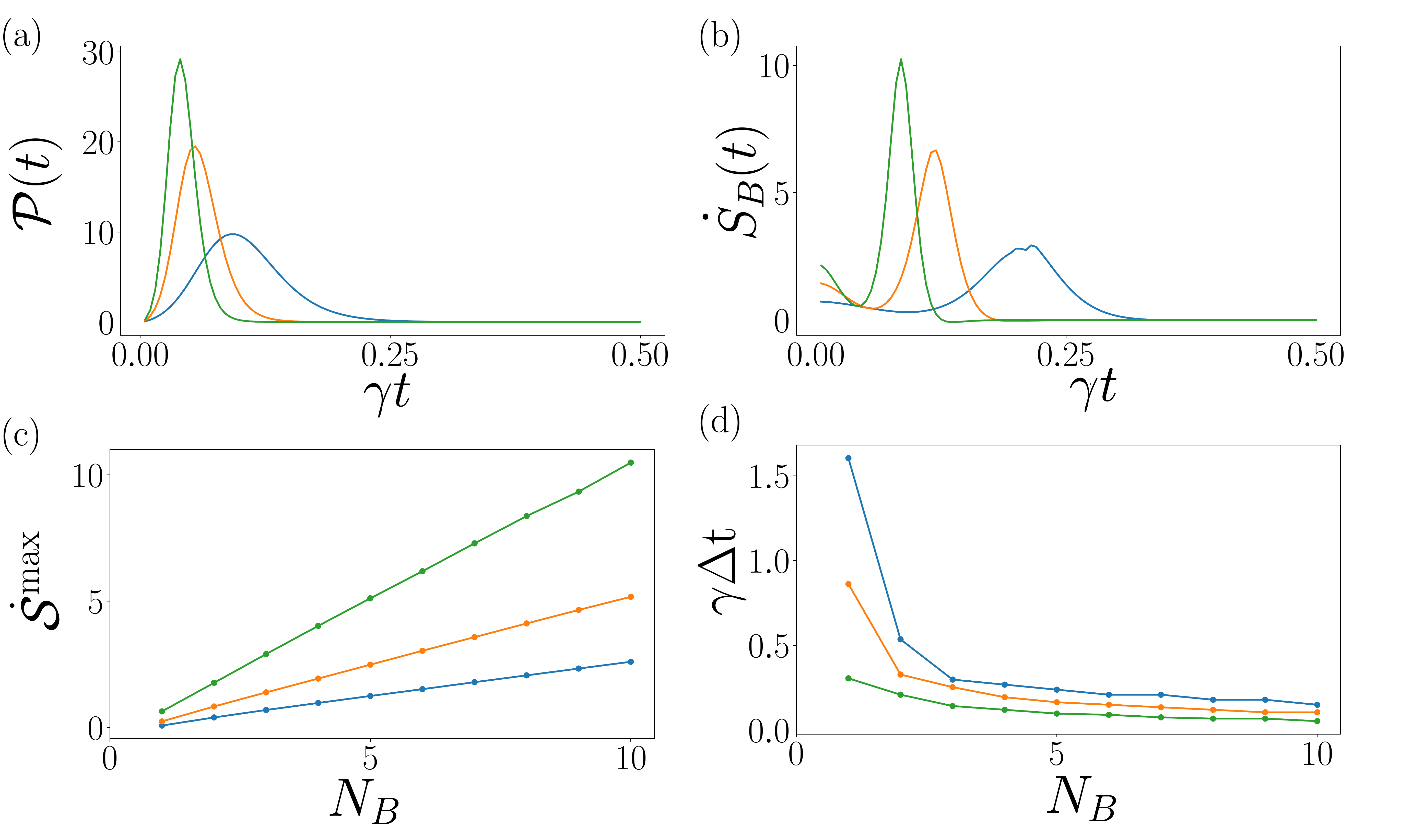}
	\caption{\textit{Entanglement rate and power density.} (a) Power density $\mathcal{P}_B(t)$. (b) Entanglement rate $\dot{S}_B(t)$. Comparing (a) and (b) shows that periods of non-zero $\mathcal{P}_B(t)$ approximately corresponds to periods of non-zero $\dot{S}_B(t)$. (c) Local maximum entanglement rate $\dot{S}_B^\mathrm{max}$ linearly scales with $N_B$. (d) The lag time between $\mathcal{P}^\mathrm{max}$ and  $\dot{S}_B^\mathrm{max}$ decreases with $N_B$. \textit{Parameters}: (a),(b) R = 50, $N_B$ = 1 (blue), 2 (orange), 3 (green). (c),(d) R = 3 (blue), 5 (orange), 10 (green).}
	\label{fig:4}
\end{figure}

If energy correlates with entanglement, then it follows that power should correlate with entanglement rate. In Fig.~\ref{fig:4}(a) and (b) we plot $\mathcal{P}_B(t)$ and $\dot{S}_B(t)$ for $N_B=1,2,3$ at $R=50$. Periods of non-zero $\mathcal{P}_B(t)$  corresponds to periods of non-zero $\dot{S}_B(t)$. In Fig.~\ref{fig:4}(c) we plot the local maximum entanglement rate $\dot{S}_B^\mathrm{max}$, for various $R$ (when there are more than one local maxima, such is the case for $N_B=1$, we choose the largest value). The plot shows that $\dot{S}_B^\mathrm{max}$ linearly scales with $N_B$. As $\mathcal{P}^\mathrm{max}$ also linearly scales with $N_B$, $\dot{S}_B^\mathrm{max}$ and $\mathcal{P}^\mathrm{max}$ are positively correlated. Interestingly, $\mathcal{P}^\mathrm{max}$ and  $\dot{S}_B^\mathrm{max}$ do not occur at the same time: $\dot{S}_B^\mathrm{max}$ lags $\mathcal{P}^\mathrm{max}$ by $\gamma\Delta t$. Fig.~\ref{fig:4}(d) plots this lag time; it shows that the lag time decreases with increasing $N_B$ or $R$. In the large $N_B$ or $R$ limit, the lag time vanishes.

Another important feature revealed by the plots is that $\dot{S}_B^\mathrm{max}$ increases with $R$, whilst $S^\mathrm{ss}_B$ decreases. This correlates with the observation that $\mathcal{P}^\mathrm{max}_B$ superextensively increase with $R$ [Fig.~\ref{fig:2}(d)], whilst the superextensivity of  $\mathcal{E}^\mathrm{ss}_B$ diminishes with $R$ [Fig.~\ref{fig:2}(b)]. These correlations provide further evidence that entanglement underpins the superextensive properties of the battery.

In unitary systems with global interaction, it has been shown that entangled states reduce the number of operations required to reach a passive state, thereby increasing power~\cite{hovhannisyan13,binder15}; the rate at which entangled states are generated does not seem to play a part. Here we show something different. In our non-unitary system with local interactions, we show that for a given $R$, energy is correlated with the level of entanglement, and power is related to the rate at which this entanglement is generated. This suggests a different mechanism for driving superextensive behavior with entanglement in our protocol. 

\section{Temperature.}
\label{sec:Temperature} 
 The effects of of thermal fluctuations on the battery provides a rich area of investigation; here we show some interesting properties. Let us begin by considering two spins at non-zero temperature. From Eq.~(\ref{eq:rho_ss_T})we can determine the spin expectation values of the charger and battery for non-zero temperature,
\begin{equation}
\Exp{J_C^z}=\Exp{J_B^z}=-\frac{2\bar{n}+1}{12\bar{n}(\bar{n}+1)+4}\hbar~.
\label{eq:energyTemp}
\end{equation}
At high temperature $\Lim{T \rightarrow \infty} \Exp{J_i^z} = 0~$, meaning thermal fluctuations dominate so that spins are equally as likely to found in the spin-up as spin-down state. At low temperature the battery obtains its energy primarily from the charger, but as the temperature increases the energy source shifts from the charger to the reservoir. 

\begin{figure}[h]
	\includegraphics[width=\columnwidth]{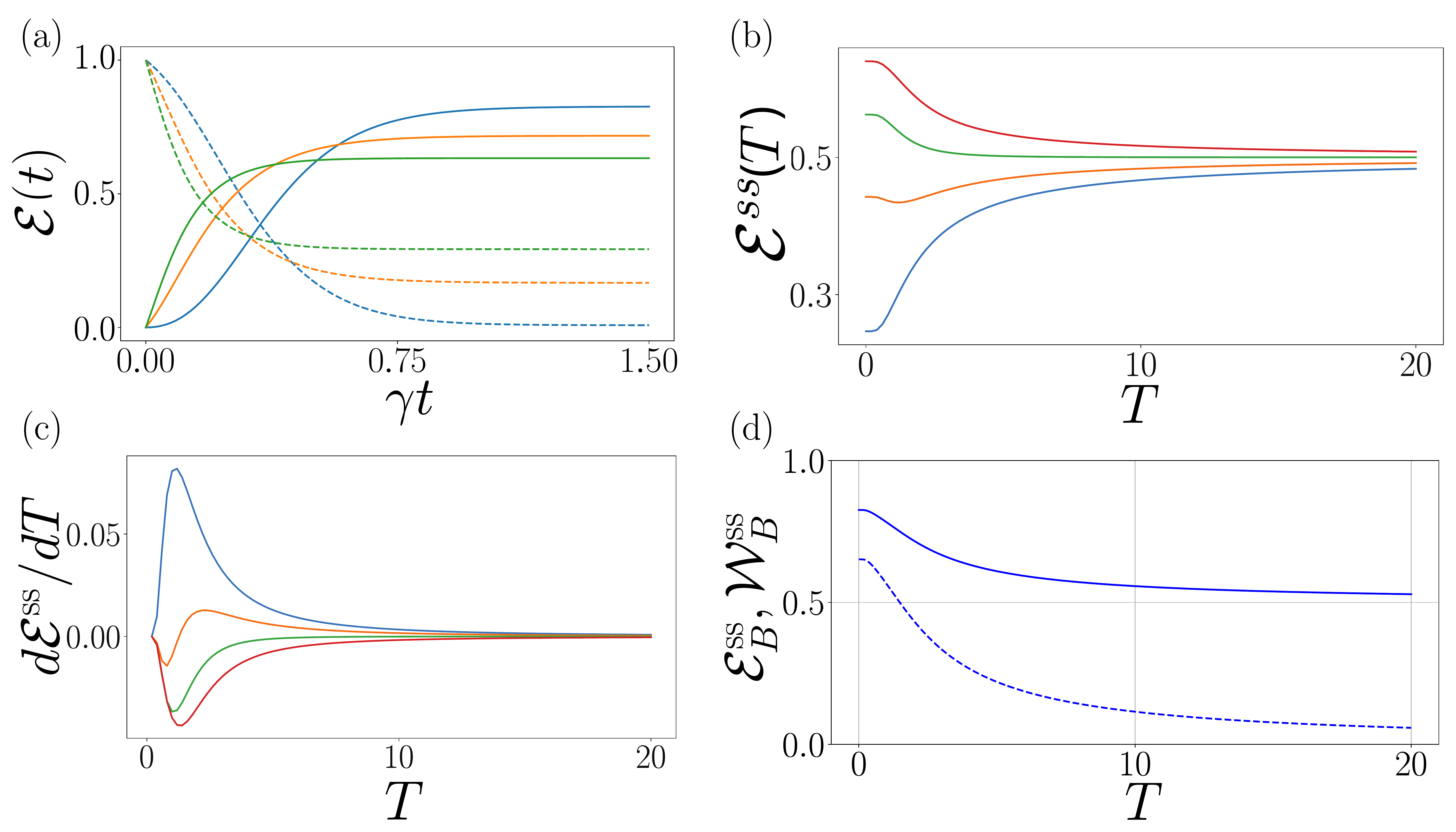}
	\caption{\textit{Charger and battery performance at non-zero temperature.} (a) Energy density of the charger $\mathcal{E}_C(t)$ and battery $\mathcal{E}_B(t)$ during the charging process for various temperatures. As the temperature increases, less energy is transferred from the charger to that battery.  (b) Steady state energy density $\mathcal{E}^\mathrm{SS}_B(T)$ for various $R$. All states converge in the thermodynamic limit to $\mathcal{E}_B^\mathrm{SS} = \frac12$~. (c) A plot of the rate of change in the steady state energy density against temperature $d\mathcal{E}_B^\mathrm{SS}/dT$. There is a decline in $\mathcal{E}_B^\mathrm{SS}$ at low temperatures, as thermal fluctuation destroy the dark states. This is followed by a deceleration in the loss of energy as the system thermalizes. The exception is for $R=1$, where the infusion of energy from thermal reservoir more than compensates for the loss of energy from the destruction of the dark state. (d) A plot of $\mathcal{E}_B^\text{SS}$ (solid) and $\mathcal{W}_B^\text{SS}$ (dotted). At high temperatures, $\mathcal{W}_B^\text{SS}\rightarrow0$ as $\mathcal{E}_B^\text{SS}\rightarrow\frac12$. \textit{Parameters}: (a) T = 0 (blue), 2 (orange), 4 (green), R = 10, $N_B$ = 1. (b),(c) R = 1 (blue), 2 (orange), 3 (green), 4 (red), $N_B$ = 1. (d) $R=4$. The vertical axes are in units of $\hbar\omega$.}
	\label{fig:5}
\end{figure}

This behavior is generalized to various $R$ as shown in Fig.~\ref{fig:5}, where we we plot $\mathcal{E}_i(t)$ for increasing $T$. Fig.~\ref{fig:5}(a) shows that as the temperature increases, less energy is transferred from the charger to the battery. In Fig.~\ref{fig:5}(b) we plot $\mathcal{E}^\mathrm{SS}_B(T)$. It shows that all states converge to  $\Lim{T \rightarrow \infty} \mathcal{E}_B^\mathrm{SS} = \frac12~$, as the system thermalizes. For states where $\mathcal{E}^\mathrm{SS}_B<\frac12$ at $T=0$, thermal fluctuations increases the battery capacity. Conversely for states where $\mathcal{E}^\mathrm{SS}_B>\frac12$ at $T=0$, thermal fluctuations decreases battery capacity. However there is a trade-off between the infusion of energy from the reservoir, and the destruction of dark states caused by thermal fluctuations. As shown in Fig.~\ref{fig:A1}, for $R=1$, $\mathcal{E}_B^\mathrm{SS}$ increases with temperature, as the infusion of energy from the reservoir more than compensates for the loss of energy from the destruction of dark states. Conversely, for $R>3$, $\mathcal{E}_B^\mathrm{SS}$ decreases with temperature, with the greatest decline occurring at low temperature, as the infusion of energy from the reservoir cannot compensate for the destruction of dark states.  $R=2$ is an interesting intermediary case, as $\mathcal{E}_B^\mathrm{SS}$ can both increase or decrease, depending on the temperature. 

As previously mentioned, a non-zero temperature lowers the upper bound on extractable work. This is reflected in Fig.~\ref{fig:5}(d) which shows $\mathcal{W}_B^\text{SS}\rightarrow0$ as $\mathcal{E}_B^\text{SS}\rightarrow\frac12$, since one would not expect there to be any extractable work under unitary transformations as the system thermalizes.

\section{Implementation and applications}
\label{sec:Implementation} Our protocol can be implemented with atomic or artificial two-level systems, including superconducting qubits, semiconductor quantum dots, ultracold atoms, trapped ions, and nitrogen-vacancy (NV) centers. We propose that experimental verification should be conducted in two regimes. Our protocol should be investigated deep in the quantum regime with few spins and at low temperature, but with a high level of control and measurement. As such, superconducting qubits coupled to a broad band resonator, which acts as the reservoir, would be suitable~\cite{gu17}. However, this platform typically is limited to few qubits.

Although QB capacity on small energy scales may find application in quantum technologies, verifying the ability to scale up capacity is important for wider adoption. Therefore, we propose that the protocol should also be investigated in the semi-classical regime with many spins and high temperature. NV centers coupled to a broad band resonator, would be a suitable platform to achieve this. Large coherent ensembles of NV-center spins ($>10^{16}$) coupled to superconducting circuits have been used to demonstrate the collective behavior of superradiance~\cite{angerer18}, and the coherent coupling between two macroscopically separated spin ensembles has also been realized~\cite{astner17}.

Because QBs utilize quantum properties, they should find applications in other quantum technologies, such as quantum computing, communication, sensing. As these technologies are underpinned by the quantum storage and transfer of energy, the applications of QB devices or principles to these technologies has the potential to improve their functionality, possibly opening new fields of investigation. For example, superextensive charging may increase quantum computation power, enhance quantum capacitor capabilities, and QB principles could advance quantum sensing devices.

Whether QBs can replace conventional batteries is ultimately a question of scalability. Nevertheless, QB devices and principles will need to find novel ways to interface with conventional technologies. An example of how quantum technology can find novel application in classical devices, is provided by the quantum dot solar cell. Here the tunable band gap of quantum dots replaces the fixed band gap of conventional bulk materials such as silicon, copper indium gallium selenide (CIGS) or cadmium telluride (CdTe). QB principles have significant  potential to find applications in solar cells, as its superextensive charging property may be utilized to superabsorb light.

\section{Conclusion}
\label{sec:Conclusion}
Our protocol is major step towards the experimental realization of a QB that achieves superextensive capacity and charging: it uses only local interactions, and is intrinsically stable in an open system - two critical features for practical applications. This rich protocol opens the way for further theoretical investigation, including a deeper understanding of the correlation between entanglement rate and power.

\section*{Acknowledgments}
\label{sec:Acknowledgements}
We thank Jared Cole, Sergi Julia-Farre, Andreas Angerer, Johannes Majer and Kae Nemoto for valuable discussions.  JQQ acknowledges the Ramsay fellowship and the Centre for Nanoscale BioPhotonics Family Friendly Fund, for financial support of this work. WJM acknowledges partial support for this work from a Japanese MEXT Grant-in-Aid for Scientific Research(A) KAKENHI Grant No. 19H00662 and the MEXT Quantum Leap Flagship Program (MEXT Q-LEAP) Grant No. JP- MXS0118069605.

\section*{Appendix: Derivation of the steady state of two spins in a thermal reservoir}
\label{sec:Derivation of the steady state of two spins in a thermal reservoir}

Here we derive the steady state of two spins in a thermal reservoir, which gives Eq.~(3) in the main text. We begin by defining the following spin basis:
\begin{align}
\ket{\psi_\uparrow}&\equiv \ket{\frac12}_C\ket{\frac12}_A \equiv \ket{1}\\
\ket{\psi_+}&\equiv\frac{\ket{\frac12}_C\ket{{-\frac12}}_A+\ket{{-\frac12}}_C\ket{\frac12}_A}{\sqrt{2}}\equiv \ket{2}\\
\ket{\psi_-}&\equiv\frac{\ket{\frac12}_C\ket{{-\frac12}}_A-\ket{{-\frac12}}_C\ket{\frac12}_A}{\sqrt{2}}\equiv \ket{3}\\
\ket{\psi_\downarrow}&\equiv \ket{{-\frac12}}_C\ket{{-\frac12}}_A\equiv \ket{4}
\end{align}
From the Linblad master equation, we write down the equations of motion for the elements of the Hermitian density matrix in the spin basis defined above [$\rho_{ij}\equiv\Exp{i|\rho|j}$]:
\begin{align}
\dot{\rho}_{11}&=-2\gamma(\bar{n}+1)\rho_{11}+2\gamma \bar{n}\rho_{22}\\
\dot{\rho}_{22}&=2\gamma(\bar{n}+1)\rho_{11}-2\gamma (2\bar{n}+1)\rho_{22}+2\gamma \bar{n}\rho_{44}\\
\dot{\rho}_{33}&=0\\
\dot{\rho}_{44}&=2\gamma(\bar{n}+1)\rho_{22}-2\gamma \bar{n}\rho_{44}\\
\dot{\rho}_{12}&=-[\gamma(3\bar{n}+2)-i\omega]\rho_{12}\\
\dot{\rho}_{13}&=-(\gamma \bar{n}-i\omega)\rho_{13}\\
\dot{\rho}_{14}&=-[\gamma(2\bar{n}+1)-i2\omega]\rho_{14}\\
\dot{\rho}_{23}&=-(\gamma(2\bar{n}+1)\rho_{23}\\
\dot{\rho}_{24}&=-[\gamma(3\bar{n}+1)-i\omega]\rho_{24}\\
\dot{\rho}_{34}&=-(\gamma \bar{n}-i\omega)\rho_{34}
\end{align}

Solving these equations one finds that in the steady state, the off-diagonal terms vanish, leaving only the diagonal terms given by:
\begin{align}
\rho_{11}^\mathrm{ss}&=\frac{\bar{n}^2[1-\rho_{33}(0)]}{1+3\bar{n}(\bar{n}+1)}\label{eq:rho11}\\
\rho_{22}^\mathrm{ss}&=\frac{\bar{n}(\bar{n}+1)[1-\rho_{33}(0)]}{1+3\bar{n}(\bar{n}+1)}\label{eq:rho22}\\
\rho_{33}^\mathrm{ss}&=\rho_{33}(0)\label{eq:rho33}\\
\rho_{44}^\mathrm{ss}&=\frac{(\bar{n}+1)^2[1-\rho_{33}(0)]}{1+3\bar{n}(\bar{n}+1)}\label{eq:rho44}
\end{align}

The initial state in the spin basis has non-zero elements: $\rho_{22}(0)=\rho_{23}(0)=\rho_{32}(0)=\rho_{33}(0)=1/2$~. It is then straightforward to show the steady state density matrix has the form
\begin{equation}
\begin{split}
\rho^\text{ss}&=\frac12[\bar{n}^2\ket{\psi_\uparrow}\bra{\psi_\uparrow}+\bar{n}(\bar{n}+1)\ket{\psi_+}\bra{\psi_+}\\
&\quad+\ket{\psi_-}\bra{\psi_-}+(\bar{n}+1)^2\ket{\psi_\downarrow}\bra{\psi_\downarrow}]\\
&\quad/[1+3\bar{n}(\bar{n}+1)]~.
\end{split}
\label{eq:rho_ss_T}
\end{equation}
Eq.~(3) of the main text is obtained by setting $\bar{n}=0$. From the density matrix one can get the spin expectation values through $\Exp{J_i^z}=\mathrm{Tr}(\rho_iJ_i^z)$.

\bibliographystyle{abbrvunsrt}

\bibliography{ntqb}

\end{document}